\documentclass[prd,aps,twocolumn,showpacs]{revtex4}
\usepackage{amsfonts}
\usepackage{amsmath}
\usepackage{graphicx}
\begin{document}
\title{Particle Swarm Optimization and gravitational wave data analysis:
 Performance on a binary inspiral testbed}
\author{Yan Wang}
\affiliation{Department of Astronomy, Nanjing University, Nanjing, 210093, China}
\altaffiliation[Current address:]{ Albert Einstein Institute, Callinstr. 38, 30167 Hannover, Germany }
\author{Soumya D. Mohanty}
\affiliation{Center for Gravitational Wave Astronomy, Dept. of Physics and Astronomy, The University of
Texas at Brownsville, 80 Fort Brown, Brownsville, TX 78520, U.S.A.}
\email[Corresponding author:]{mohanty@phys.utb.edu}
\begin{abstract}
The detection and estimation of gravitational wave (GW) signals belonging to a parameterized family of waveforms
 requires, in general,  the numerical maximization of a data-dependent function of
the signal parameters.
Due to noise in the data, the function to be
maximized is often highly multi-modal with numerous local maxima.  Searching for the global maximum then becomes computationally
expensive, which in turn can limit the scientific scope of the search. Stochastic
optimization is one possible approach to
reducing computational costs in such applications.
 We report results from
a first investigation of the Particle Swarm Optimization (PSO) method in this context.
The method is applied to a testbed motivated by the  problem of
detection and estimation of a binary inspiral signal. Our results show that PSO works well
in the presence of high multi-modality, making it a viable candidate method for further applications in  GW data analysis.
\end{abstract}
\maketitle
%%%%%%%%%%%%%%%%%%%%%%%%%%%%%%%%%%%%%%%%%%%%%%%%%%%%%%%%%%
\section{Introduction}
\label{introduction}
The detection and estimation of a gravitational wave (GW) signal belonging to a parameterized
family  of waveforms
requires, in general, the numerical maximization of some data-dependent function
 over the space of the signal parameters. For example,
in the {\em matched filtering}~\cite{helstrom,wainstein+zubakov} method, which is the focus of this paper,
the function to be maximized  is
 a suitably defined inner product between the data and parameterized signal waveforms.
The global
maximum of this function serves as a detection statistic.
A point estimate of the signal parameters is furnished
by the  location of the global maximum in parameter space.

The presence of noise in the output of GW detectors leads to  a large number of
local maxima in this function that are distributed randomly in parameter space. The search for
the global maximum in this forest of local maxima then becomes a computationally
expensive task. This can affect the sensitivity of a search by limiting either the
volume that is searched in parameter space or the integration
length of data required for accumulating
sufficient signal-to-noise ratio (SNR), or both.
The computational
efficiency of the search for the global maximum is, thus, an important issue in GW data analysis.
The various search strategies proposed in the GW literature so far
 can be broadly divided into those based on sampling the function on predetermined
 grids of points in parameter space~\cite{owen96,owen99,mohanty+dhurandhar:96}, and those that use
 stochastic optimization methods (e.g.,~\cite{christensen04,cornish+crowder:2005,wickham+etal:2006}).

In the class of grid-based methods, significant
savings in computational costs have been demonstrated with a hierarchy of grids~\cite{mohanty+dhurandhar:96,mohanty98,sengupta+etal}.
 A nice feature of grid-based methods is that they
are easy to characterize statistically and, hence, design variables of the algorithm, such as the
spacing of points, can be fixed systematically.

Stochastic methods do not use
 pre-determined grids but employ some form of
 random walk through the parameter space.
The probabilistic rules of the random walk are tuned to maximize the chances of
its terminating close to the global maximum. There are many algorithms that fall
under the class of stochastic methods, a hybrid of simulated annealing and Metropolis-Hastings
Markov Chain Monte Carlo (MCMC) being the most widely explored in GW
data analysis~\cite{christensen04,cornish+crowder:2005,wickham+etal:2006}.

Since the number of points in a grid grows exponentially with the dimensionality
of the parameter space, stochastic methods tend to outperform grid-based
ones with an increase in the number of signal parameters.
It is worth noting here that stochastic methods in GW data analysis
incur the additional computational cost of generating signal waveforms
on the fly. In grid-based methods, on the other hand, waveforms can be computed
and stored in advance of processing the data. Stochastic methods can, therefore, lose their advantage if
the computational cost of generating waveforms becomes too high.

The performance of a stochastic method may be sensitive to the values to which its design variables
are tuned. 
Since the tuning
 is usually done on simulated data, it is not clear how robust current stochastic methods are
 against features of real data such as non-stationarity and non-Gaussianity. 
Additionally, the number of design variables that require careful tuning is fairly large for some of the methods. In
such cases, tuning becomes more of an art than a well defined procedure and this may also
affect robustness. In some methods, prior information is used about generic features of the function 
to be maximized. This may not be reliable if the assumptions behind the prior information, such
as a particular noise model, become 
invalid.
To properly address
issues such as these it is important 
that a wide variety of stochastic methods be explored in GW data analysis .

Particle Swarm Optimization (PSO)~\cite{kennedy95}, first proposed by Kennedy and Eberhardt in 1995,
is a stochastic method
that has been garnering a lot of attention recently in many application areas~\cite{proc:SIS09}.
An attractive feature of PSO is that,
 in its basic form, it has a small number of design variables.
On standard testbeds, PSO has been found
to have comparable or superior performance to other well known methods such as MCMC.

 This paper presents the first application of PSO to GW data analysis. We pose the following specific questions:
\begin{enumerate}
\item Is PSO a viable method when applied to a function that is highly multi-modal and essentially stochastic in nature? This is the 
typical case in GW data analysis.
\item How many design variables are there in PSO, and how many of them need to be tuned well?
\item Can the tuning of these variables be done without  requiring prior information about features of the function, thus increasing the
 robustness of the method?
\item What is the computational cost of the method and what are the most important
technical improvements required for the future?
\end{enumerate}
To answer these questions in the most direct and reliable manner, we construct a testbed based on the
well understood task of detecting and
estimating binary inspiral signals in data
from a single ground-based detector. This problem involves low-dimensionality but
offers the more serious challenge of high multi-modality.
To keep the focus
on the latter, a simplification
is made regarding the shape of the search region such that it admits
unphysical waveforms. Thus, the implementation of PSO presented here is
not directly applicable to binary inspiral searches at present. The required technical refinements are
discussed in the paper.
In addition, a novel and systematic tuning procedure is introduced that is
based on data containing only noise. This  procedure may be useful for other stochastic methods also.

The rest of the paper is organized as follows. Sec.~\ref{testbed} describes the testbed and
Sec.~\ref{PSOdescription} describes the PSO method. We explain our procedure for tuning the
design variables of PSO in Sec.~\ref{tuning}.  Sec.~\ref{results} then presents results from numerical simulations.
 Our conclusions and pointers to future work are presented in Sec~\ref{conclusions}.

%%%%
\subsection*{Notation}
\begin{description}
\item{${\bf x}$, ${\bf y}$, etc}: A time series with a finite number, $N$, of samples. The $k^{\rm th}$ sample, $0\leq k\leq N-1$, is denoted by $x[k]$.
%%%%%%%%%
\item{$\delta_s$, $T$}: The sampling interval and the duration of ${\bf x}$ respectively.
The number of samples in ${\bf x}$ is  $N=[T/\delta_s]$, where the square brackets denote
 truncation to the nearest integer.
%%%%%%%%%
\item{$\Theta$} : The set of parameters describing a family of signals.
%%%%%%%%%
\item{${\bf s}(\Theta)$}:
The time series of the signal corresponding to parameter values
  $\Theta$. The $k^{\rm th}$ sample of ${\bf s}(\Theta)$ is denoted by $s[k;\Theta]$. In our case, signals have a well-defined start and stop time and the interval between them may be less than $T$. However, ${\bf s}(\Theta)$ still consists of $N$ samples with the samples outside the interval enclosed by the start and stop times set to 0 (zero-padding).
%%%%%%%%%%
\item{$\widetilde{\bf x}$}: The Discrete Fourier Transform (DFT) of ${\bf x}$. The DFT value at the
 frequency $k/T$, $k=0,1,\ldots,[N/2+1]$, is denoted as $\widetilde{x}[k]$. The DFT of ${\bf s}(\Theta)$ is denoted by
 $\widetilde{\bf s}(\Theta)$ and its value at the $k^{\rm th}$ frequency by $\widetilde{s}[k;\Theta]$.
%%%%%%%%%
\item{$\langle {\bf x},{\bf y}\rangle$}: The time series introduced above are elements of
 $\mathbb{R}^N$, the vector space of real $N$-tuples. From the point of view of
 detection and estimation of a signal in data with additive stationary noise, a natural inner
product can be introduced on this vector space,
\begin{equation}
\langle {\bf x},{\bf y}\rangle = 4\mathcal{R}\left(\sum_{k=0}^{[N/2+1]}{\frac{\widetilde{x}^*[k]\widetilde{y}[k]}{S_n[k]T}}\right)
\end{equation}
where $S_n[k]$ is the one-sided Power Spectral Density (PSD) of the noise.
\item{$\|{\bf y}\|$}: the norm on $\mathbb{R}^N$,
\begin{equation}
\|{\bf y}\|^2 = \langle {\bf y},{\bf y}\rangle\;,
\end{equation}
induced by the inner product defined above.
%%%
The signal to noise ratio (SNR) of a signal ${\bf s}(\Theta)$ is defined as $\|{\bf s}(\Theta)\|$.
\end{description}

%%%%%%%%%%%%%%%%%%%%%%%%%%%%%%%%%%%%%%%%%%%%%%%%%%%%%%%%
\section{Test bed}
\label{testbed}
In this section, we describe the testbed to which PSO is applied. The testbed is constituted by
the noise model, signal family and
the function to be maximized.
%%%%%%
\subsection{Noise model}
A GW signal incident on an interferometric ground-based detector produces a
difference in the lengths of its two arms. After calibrating out the common arm length and
the transfer function of the detector, the
data, ${\bf x}$, contains the measured GW-induced strain added to
instrumental and environmental noise ${\bf n}$.
 Thus, ${\bf x}={\bf n}$ when no GW
signal is present, and ${\bf x}={\bf s}(\Theta)+{\bf n}$ when there is.
In our simulations,  ${\bf n}$ is a realization of a stationary, Gaussian noise process
 with a PSD, $S_n[k]$,
 that matches the initial LIGO~\cite{ligo} design sensitivity curve in shape~\cite{ifo_sensitivity}.

\subsection{Signal waveforms}
\label{waveform} We use the signal family associated with a
non-spinning inspiraling binary system, computed up to the second
post-Newtonian (2PN) order~\cite{blanchet+etal}. This system
consists of two non-spinning compact stars (Neutron Stars or Black
Holes)   losing orbital binding energy
through GW emission. Members of this signal family have {\em chirp}
waveforms with monotonically increasing
 instantaneous amplitude and frequency.

For the case of a single detector, the parameters specifying the 2PN
signal waveforms can be grouped into two sets. The first set is that of the
 {\em chirp-time}~\cite{bss:chirptimes} parameters, $\{\tau_a\}$, $a=0,1,1.5,2$, that are constructed out of the masses of the two
components of the binary. Expressions for the chirp-time parameters are provided in
Appendix~\ref{details_signal}. The second set consists of the {\em time-of-arrival}, $t_a$, the {\em initial phase}, $\Phi_a$
and the amplitude, $\mathcal{A}$. Interferometric ground-based detectors have a sharp rise in seismic noise below
 some frequency $f_a$ ( $=40$~Hz for inital LIGO). The chirp signal from a binary inspiral is essentially unobservable when its
 instantaneous frequency is below $f_a$. The time at
which the signal becomes visible is $t_a$ and the corresponding
 instantaneous phase of the signal is $\Phi_a$.

Since all the four chirp-times depend on the masses of the two compact stars, only two of them are independent.
We choose $\tau_0$ and $\tau_{1.5}$ as the two independent chirp-time parameters.
Thus, the set of signal parameters
is $\Theta = \{ \mathcal{A}, \Phi_a, t_a, \theta =\{\tau_0,\tau_{1.5}\}\}$.

As discussed in Sec.~\ref{introduction}, the computational
cost of generating waveforms on the fly is important for stochastic methods like PSO.
The 2PN signal family is amenable to
a fast implementation because a sufficiently accurate analytical form  exists for the
Fourier transform of these waveforms~\cite{droz+etal},
\begin{widetext}
\begin{eqnarray}
\widetilde{s}[k;\Theta]&=&\left\{
\begin{array}{ll}
0\;, & k\leq[f_a T]\;,\\
\mathcal{A}\mathcal{N}
f^{-7/6}\exp\left[-2\pi i f t_a+i\Phi_a-i\psi(f;\theta)+i\frac{\pi}{4}\right]\;,& [f_aT]<k\leq [f_c T]\;,\\
0\;, & k > [f_c T]
\end{array}\right.
\label{stat_phase_eqn}
\end{eqnarray}
\end{widetext}
where, the lower cutoff frequency $f_a$ was explained above and
the upper cutoff frequency $f_c$ follows from the termination of the inspiral waveform when
the binary reaches its last stable orbit.  For our testbed, we set $f_c=700$~Hz although in a real search it depends on the
mass of the binary system.
  The expression for  $\psi(f;\theta)$ is given in Appendix~\ref{details_signal}.
The normalization
constant $\mathcal{N}$ is defined such that,
$ \|{\bf s}(\{\mathcal{A}=1,\Phi_a,t_a,\theta\})\|^2 = 1$.
It follows  that $\mathcal{A}$ is the SNR of the signal.

Later on, we use the fact that although not all values of $\theta$ correspond to valid binary mass components,
  Eq.~\ref{stat_phase_eqn} can still be used to generate perfectly normal
waveforms. These waveforms are also chirps but their phase evolution does not correspond to any
physical binary system.

%%%%%%%%%%%%%%%%%%%
\subsection{Fitness function}
\label{detection_statistic}
The function to be maximized is,
\begin{eqnarray}
\Lambda(t_a,\theta|{\bf x}) &=&
 \left[\langle {\bf q}_0(t_a,\theta),{\bf x}\rangle^2+\right.\nonumber\\
&&\left. \langle {\bf q}_{\pi/2}(t_a,\theta),{\bf x}\rangle^2\right]^{1/2}\;,
\label{mf_functional_actual}\\
{\bf q}_{\phi}(t_a,\theta)&=& {\bf s}(\{\mathcal{A}=1,\Phi_a=\phi,t_a,\theta\})\;.
\end{eqnarray}
This function is obtained by maximizing the
log-likelihood, $\langle {\bf x},{\bf s}(\Theta)\rangle-(1/2)\|{\bf s}(\Theta)\|^2$ analytically over $\mathcal{A}$ and $\Phi_a$.

For given $\theta$, the evaluation of $ \langle {\bf q}_\phi(t_a,\theta),{\bf x}\rangle$ over $t_a=m\delta_s$, $m=0,1,\ldots,N-1$,
is a cross-correlation operation that can be computed efficiently using the Fast Fourier Transform (FFT).
Thus, the function that is
 maximized using PSO is,
\begin{equation}
\lambda(\theta|{\bf x}) = \max_{t_a}\Lambda(t_a,\theta|{\bf x})\;.
\end{equation}
 In the remainder of the paper, $\lambda(\theta|{\bf x})$ will be called the
{\em fitness function} in keeping with the standard terminology used in much of the literature on
stochastic methods.

\begin{figure}
\includegraphics[scale=0.6]{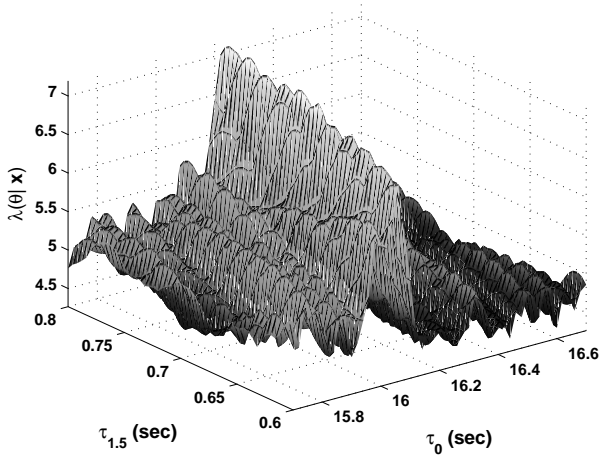}
\caption{\label{gridsnr}
A realization of the fitness function for the binary inspiral testbed. The data contains a signal with
an SNR=8.0. In the absence of noise, the fitness function has  only one extremum at the location
identified by the chirp-times of the signal. The presence of noise leads to a forest of local maxima.
}
\end{figure}

The presence of noise in ${\bf x}$ makes the fitness function highly multi-modal as shown in Fig~\ref{gridsnr}.
The large number of local maxima with random locations and sizes poses a strong challenge to stochastic
methods. When the noise is stationary and Gaussian, and the signal present in the data is from the waveform family
that one is searching for, certain characteristic features are present in the fitness function. For example, the shape
of the peak in Fig.~\ref{gridsnr} is elongated on the average along a predictable direction. MCMC methods in the GW literature
use this type of prior information about the fitness function in tuning the design variables~\cite{cornish+crowder:2005}.

 %%%%%%%%%%%%%%%%%%%%%%%%%%%%%%%%%%%%%%%%%%%%%%%%%%%%%%%%
%%%%%%%%%%%%%%%%%%%%%%%%%%%%%%%%%%%%%%%%%%%%%%%%%%%%%%%%
\section{Particle Swarm Optimization}
\label{PSOdescription}

The
PSO algorithm is first described in terms of a general fitness function $\lambda(\theta)$, over
some parameter set $\theta$. Later, we specialize the discussion to the case of the binary inspiral testbed.

\subsection{The PSO Algorithm}
\label{PSOalgorithm}

Let $\theta = \{\theta_1,\theta_2,\ldots,\theta_D\}$ denote
a point in $\mathbb{R}^D$, and $\lambda(\theta)$ be the fitness function.
The essential idea behind PSO is to compute $\lambda(\theta)$ simultaneously
at several locations and use these samples to influence
the locations for computing the next set of samples. This
process continues iteratively until some stopping rule is satisfied.
The process can be visualized by treating the sample locations
 as a {\em swarm} of {\em particles} that moves in $\mathbb{R}^D$, hence
the name of the algorithm.  A precise description now follows.

Let the
coordinates in $\mathbb{R}^D$ of
the $i^{\rm th}$ particle in a swarm of $N_p$ particles be
 $Q_i[k]$ at the $k^{\rm th}$ step in the search $(k=0,1,\ldots$).
Associated with this particle is a {\em velocity} vector $V_i[k]$
that determines $Q_i[k+1]$,
\begin{eqnarray}
Q_i[k+1]&=&Q_i[k]+V_i[k]\label{PSO_coord}\;.
\end{eqnarray}
The PSO algorithm is usually started with randomly chosen particle locations and velocities. In our
implementation, we position the particles initially on a regular grid while the initial velocities are
kept random.

 Let the maximum value of $\lambda(\theta)$ found by the $i^{\rm th}$ particle over $k$ steps  be $R_i(k)$
and the location of $R_i(k)$, called the particle's best location
{\em pbest},  be
 $P_i[k]$. Thus,
\begin{equation}
R_i(k)=\lambda(P_i[k])\geq \lambda(Q_i[j]);\;  j\leq k\;.
\end{equation}
 Let the maximum
 over $\{R_i(k)\}$, $i=1,\ldots,N_p$,  be $R_g(k)$ and
its location, called
the global best location {\em gbest} be $P_g[k]$,
\begin{equation}
R_g(k)=\lambda(P_g[k])\geq \lambda(P_j[k]);\;\forall j\;.
\end{equation}
At any step, there is always one particle in the swarm whose {\em pbest} is also the {\em gbest}.
We call this particle the {\em best particle} at step $k$.
Note that both {\em pbest} and {\em gbest} are locations found over
the entire past history of the motion of the particles. They need not necessarily change at every step.

 The velocity for the $i^{\rm th}$ particle at the next step, $k+1$, is determined
by the dynamical equation,
\begin{eqnarray}
V_i[k+1] &=& wV_i[k]+c_1\chi_1(P_i[k]-Q_i[k])+\nonumber\\
&&c_2\chi_2(P_g[k]-Q_i[k])\;,\label{PSO_vel}
\end{eqnarray}
where  $w$, which
can depend on $k$, is
 called the {\em inertia weight}, $c_1$ and $c_2$ are called
 {\em acceleration constants}, and $\chi_1$, $\chi_2$
are  random numbers drawn independently at each step
from the uniform distribution on $[0,1]$.

Finally, for any component $V_{i,m}[p]$ of the particle velocity
$V_i[p]=(V_{i,1}[p],\ldots,V_{i,D}[p])$,
\begin{equation}
V_{i,m}[p] = \left\{\begin{array}{cc}V_{{\rm max},m}\;,& V_{i,m}[p]>V_{{\rm max},m}\\
-V_{{\rm max},m}\;, & V_{i,m}[p] < - V_{{\rm max},m}\end{array}\right.\;,
\end{equation}
 where $V_{{\rm max},k}>0$, $k=1,\ldots,D$, and $V_{\rm max}=(V_{{\rm max},1},\ldots,V_{{\rm max},D})$ is called
the {\em maximum velocity}.

Like all stochastic methods, PSO involves a competition between  wide ranging exploration of the
fitness function and convergence to a best value. In order to avoid trapping by a local maximum, the method
must be able to explore other parts of the parameter space, while to find the global maximum, the method must
eventually explore a progressively smaller region around some point. The way this competition is implemented in the PSO algorithm is
seen clearly from Eqs.~\ref{PSO_vel}. The first term simply moves a particle along a straight line, while the remaining two terms are
sources of acceleration, one pulling it towards  its {\em pbest}
 and another pulling it towards {\em gbest}. The last
two effects are combined with random weights $\chi_1$ and $\chi_2$.
The random deflections and inertial motion allow a particle to explore the fitness function, while
 the attractive pulls of {\em pbest} and {\em gbest} counter this behavior. With a dynamic inertia weight that decreases
in time, the attractive pull eventually wins over. A rudimentary emulation of real biological swarming behavior is built in 
through each particle being aware of {\em gbest}.

The PSO algorithm has another interesting feature. The best
 particle, by definition, has its {\em pbest} $P_i[k]$ coincident with {\em gbest}, $P_g[k]$, making
 the terms  $P_i[k]-Q_i[k]$ and $P_g[k]-Q_i[k]$ equal for it. This particle then accelerates towards
{\em gbest} alone and only moves along a straight line through this location. This situation continues
 until  a new {\em gbest} is found.
In effect, one particle at any step shows a convergence behavior, exploring the neighborhood of the current {\em gbest},
 while the other particles continue their exploration.

%%%%%%%%%%%%%%%%%%%%%%%%%%%%
\subsection{Termination criterion}
\label{convergence}
For stochastic methods, the probability
of convergence to the global maximum is usually guaranteed
only in the asymptotic limit.
Hence, any practical implementation of a stochastic method must
include a criterion for terminating the search.
The criterion we adopt
for termination is specific to
the fitness function for the binary inspiral testbed and,
accordingly, $\theta$ now refers to the chirp time parameters.

 If the
particles in PSO continue to move over several steps but do not find a significantly different {\em gbest}, it is likely that the
current {\em gbest} lies close to the global maximum. A natural criterion
for termination then is to check if {\em gbest}
 stays confined to a small region over a predetermined number of steps.

When the data contains only a signal, ${\bf x}={\bf s}(\{\mathcal{A},\Phi_a,t_a,\theta\})$, the
fitness function is maximum at the location $\theta$ of the signal. ( ${\bf s}(\theta)\equiv {\bf s}(\{\mathcal{A},\Phi_a,t_a,\theta\})$
for brevity in the following since the other parameters do not figure in the fitness function.)
The fractional drop in the fitness function
for a small displacement $\Delta\theta = (\Delta\theta_1=\Delta\tau_0,\Delta\theta_2=\Delta\tau_{1.5})$ is given by,
\begin{eqnarray}
1-\frac{\lambda(\theta+\Delta\theta|{\bf s}(\theta))}{\lambda(\theta|{\bf s}(\theta))}
 & \simeq &\frac{-\sum_{i,j=1}^2{\cal H}_{ij}\Delta\theta_i\Delta\theta_j}{2\lambda(\theta|{\bf s}(\theta))}\;,\\
{\cal H}_{ij} & = &  \left.\frac{\partial^2\lambda(\theta^\prime|{\bf s}(\theta))}{\partial\theta_i^\prime\partial\theta_j^\prime}\right|_{\theta^\prime=\theta}\;,
\end{eqnarray}
where $\theta$ is the location of the signal. For a small fractional drop $\alpha$, therefore, we get an ellipsoidal region $\mathcal{S}_\alpha(\theta)$
 centered at $\theta$ such that $\lambda(\theta^\prime|{\bf s}(\theta))\geq (1-\alpha) \lambda(\theta|{\bf s}(\theta))$ if $\theta^\prime \in
\mathcal{S}_\alpha(\theta)$.

Now, the neighborhood of the global maximum in the presence of noise is also $S_\alpha(\theta)$ on the
average for a fractional drop $\alpha$. Therefore,
it is natural to choose the region of convergence to be
 $\mathcal{S}_\alpha(\theta)$ in general. This reduces the task of specifying the region to simply
choosing a value for $\alpha$. Following a convention widely used in the
GW literature~\cite{owen96}, we fix $\alpha = 0.03$.

Thus, we arrive at the following criterion for terminating
PSO. At each step $k$, (i) obtain the ellipsoid around {\em gbest}, that is,  $\mathcal{S}_\alpha(P_g[k])$.
 (ii) If the best location $P_g[k+1]$ falls outside $\mathcal{S}_\alpha(P_g[k])$, then
reset the region of convergence to the new best location, i.e., use $\mathcal{S}_\alpha(P_g[k+1])$.
(iii) If the region of convergence is not found to change over $N_t$ successive steps, then
terminate PSO.

The  termination criterion implies that if PSO terminates near the true global
maximum, the fitness value found will have a fractional drop less than $\alpha$. Consequently,
 it will have a performance comparable to a grid-based search in which the templates are spaced
according to the {\em minimal match} criterion~\cite{owen99} and the minimal match is $1-\alpha$. This is important
for situations where a grid-based search is infeasible as it guarantees that PSO will perform as well
or better. The probability of convergence to the global maximum must be high, however, and this is the objective of the tuning process
described later.
%%%%%%%%%%%%%%%%%%%%%%%%%%
\subsection{Search Boundary}
\label{pso_boundary_conditions}
Even with the termination criterion in place,
 the search region must be finite in order for PSO to terminate in a finite number of steps. Otherwise, the swarm may continue to find a
better {\em gbest} and the termination criterion may never be satisfied. This is
especially relevant in the case when the data has only noise.
 Thus, the PSO dynamics must be supplemented with appropriate boundary conditions.
Many approaches to this problem have been proposed, with a good summary provided in~\cite{boundaryConditions}.
 In this paper, we use the {\em invisible wall} boundary condition, but we also briefly describe some of the others below.

%%%%%%%%%%%%
\subsubsection{Types of boundary conditions}
\label{boundary_conditions}

The boundary conditions proposed in the PSO literature are as follows. (This list is taken from~\cite{boundaryConditions}
 and is by no means an
exhaustive one.)
\begin{description}
\item{\em Absorbing walls --} When a particle crosses a rectangular boundary,
the velocity component perpendicular to the boundary is zeroed. Eventually, this allows the particle to be pulled into the search domain.
\item{\em Reflecting walls --} As with the absorbing walls condition, the particle velocity is altered but instead of being zeroed, the velocity component perpendicular to the wall is reversed in sign. This throws the particle back into the search domain.
\item{\em Invisible walls --} No change is made to the dynamics of the boundary crossing
particle but $\lambda(\theta|{\bf x})$ is set to zero and it is not evaluated as long as the
particle stays outside the boundary.
\end{description}
We have tried all three boundary conditions but, like the authors of~\cite{boundaryConditions}, we
find that the invisible wall condition tends to perform better than the other two. %Another advantage of this

%%%%%%%%%%%%
\subsubsection{Search region for the testbed}
\label{physical_boundary}

The simplest search region in $\theta$ parameter space is a rectangle
$\tau_{0,min}\leq\tau_0\leq\tau_{0,max}$ and $\tau_{1.5,min}\leq \tau_{1.5}\leq \tau_{1.5,max}$.
 A part of this region, however, admits waveforms that do not correspond to a
physically valid binary system. This is due to the dependence of $\tau_0$, $\tau_{1.5}$
on the symmetric combinations of binary component masses $M$ and $\mu$,
 the total and reduced mass of the binary respectively, and the inequality $M\geq 4\mu$.
Nonetheless, as remarked in Sec.~\ref{waveform}, there is no technical problem in generating
waveforms corresponding to the unphysical chirp times and nothing strange happens to the fitness
function there. See Fig.~\ref{gridsnr}, for example, where a part of the parameter region shown is unphysical.
Since the primary utility of the binary inspiral problem in this paper is to provide a testbed
for PSO, this physical constraint is ignored.

The rectangular search region allows the coordinate transformation,
\begin{eqnarray}
x_1 &=& (\tau_0-\tau_{0,min})/(\tau_{0,max}-\tau_{0,min})\;,\\
x_2 &=& (\tau_{1.5}-\tau_{1.5,min})/(\tau_{1.5,max}-\tau_{1.5,min})\;,
\end{eqnarray}
 such that $x_i \in [0,1]$, $i=1,2$. In our codes, all PSO equations use $x_1$ and $x_2$ and the
corresponding velocity components.

We choose $\tau_{0, min}=0.94$~sec, $\tau_{0, max}=37.48$~sec, $\tau_{1.5,min}=0.234$~sec and $\tau_{1.5, max}=1.021$~sec.
With $\tau_0$ and $\tau_{1.5}$ along the horizontal and vertical axes respectively,
the upper right hand corner corresponds to binary component masses (in $M_\odot$) $m_1=1.1$ and $m_2=1.1$.
The lower left hand corner corresponds to $m_1=10.5$ and $m_2=9.7$.
%%%%%%%%%%%%%%%%%%%%%%
%%%%%%%%%%%%%%%%%%%%%%
%%%%%%%%%%%%%%%%%%%%%%
\subsection{PSO design variables}
\label{list_design}
One of the questions posed at the beginning was about the number of design variables in PSO.
In our implementation, there are a total of 9 that are listed below for reference.
\begin{description}
\item{$N_p$:} Number of particles in the swarm.
\item{$c_1,\,c_2$:} Acceleration constants.
\item{$V_{max}$:} Maximum velocity of a particle.
\item{$\alpha,\, N_t$:} The parameters used to specify the termination criterion for PSO.
 These parameters are not
part of the standard PSO algorithm.
\item{Parameters governing the inertia decay law:} The inertia weight is decreased in value as PSO progresses through a search.
The PSO literature is full of different types of decay laws but, in
general, it is known that a strictly linear decay law is not very
useful. We have developed the following decay law that has elements
of both linearity and non-linearity.  Let $w[k]$ be the value of the
inertia weight at step $k$,
\begin{equation}
w[k]=w_0-m (k-k_0)/N_t\;,
\end{equation}
where $w_0>0$ and $m > 0$. The parameter $k_0$ starts with an initial value of $k_0=0$ and
 is kept fixed as long as {\em gbest} stays within the current region of convergence. If {\em gbest}
exits the convergence region at some step $k^\prime$ without termination,
$k_0$ is set equal to $k^\prime$. Thus, the value of the inertia is reset to the starting value of $w_0$ every time
termination fails and the linear decay of the inertia starts anew.
\item{$N_{rep}$:} For given data ${\bf x}$, independent runs of PSO
 yield different  values of $\lambda(\theta|{\bf x})$ corresponding to the different fitness values at
termination.
This is unavoidable for any stochastic method. However, termination near the true global maximum
in independent runs of PSO on the same data should result in the clustering of the different values
found and their locations.
We can turn this argument around by
running PSO independently several times on the same data and using the formation of a cluster
as an indicator of successful termination in the vicinity of the global maximum. The number of independent
runs of PSO on the same data, $N_{rep}$, is also a design variable.
\end{description}

%%%%%%%%%%%%%%%%%%%%%%%%%%%%%

\section{Tuning the design variables}
\label{tuning}
For any stochastic method, convergence to the global maximum can only be quantified as a probability.
In some asymptotic limit, such as particle number $N_p\rightarrow \infty$ for PSO, this probability becomes
unity. However, this also implies an infinitely large computing cost. Thus the design variables
 must be tuned to find the best trade-off between
the probability of convergence and the associated computational cost.
We present here the procedure followed for tuning the design variables of PSO.

In contrast to the tuning procedure used for most MCMC methods in the
GW literature, our approach is not based on data containing a signal but data that is purely noise.
The latter is the worst case scenario for any stochastic method.
 However,
 good performance in the pure noise case more or less guarantees success when a signal is present.
Moreover, this approach to tuning avoids any bias due to the use of a particular set
of signals or SNRs.

The tuning procedure presented here can be used, in principle, to tune all the nine design variables
of PSO (c.f., Sec.~\ref{list_design}). However, applying the procedure to all of them is computationally too
expensive, at least for the objectives of this paper. We focus instead on two of the most important
variables for the performance of PSO, $N_p$ and $N_t$. For the rest, we either choose values commonly
used in the literature or simply pick reasonable ones based on our experience with PSO. Thus, we set :
$c_1=c_2=2$, $V_{max}=(0.5,0.5)$, $w_0=0.9$, $m=0.4$, $\alpha=0.03$ and $N_{rep}=5$.

\subsection{Criterion for optimal tuning}
\label{tuning_fom}

Measuring the probability of convergence for the pure noise case presents a practical problem.
In simulations where a large SNR signal is present, we know that the true
 maximum is most likely to be in close proximity to the location of the signal and it can be found reliably using, say, a small
area grid-based search.
For the pure noise case, however, the location is not known {\em a priori}, even approximately, and
the only reliable solution is a grid-based search over the entire search region.
However,  we avoid this solution because (i) the simulations become computationally very expensive, and
more importantly, (ii) it would fail for higher dimensional problems where grid-based searches are infeasible.

To circumvent this problem,
we invoke the argument outlined in Sec.~\ref{list_design} for using $N_{rep}$ wherein termination in the vicinity of the
global maximum is indicated by the clustering of the fitness and parameter values
 over independent runs of PSO.
One way to further confirm the association between a cluster and the global maximum
 is to increase the number of particles significantly and verify that a
cluster forms around the same location.
This is similar to what is done, for example, in the numerical solution of differential equations. To check that a given solution is
valid, the computational grid is made denser and the new solution is compared with the old one.
The above ideas can be quantified as follows, allowing an objective criterion for tuning to be developed.

Let there be a number of independent trials, in each of which  a new realization ${\bf n}$ of noise is obtained and
PSO is run $N_{rep}$ times
on ${\bf n}$.
Thus, in each trial, $N_{rep}$ values are obtained for each of the chirp-times $\tau_0$ and $\tau_{1.5}$, and the
corresponding fitness values $\lambda(\theta|{\bf n})$. We define a set of  $N_{rep}=5$
numbers to be clustered if at least 3 of
them lie in a range that is less than 30\% of the entire range of the 5 numbers. This definition
of clustering is applied to each of the three sets of $N_{rep}$ values above.
We then define,
\begin{description}
\item{Probability of clustering:} Let $P_{\tau_0}$, $P_{\tau_{1.5}}$ and $P_\lambda$ be the
fraction of trials in which clustering occurs for $\tau_0$, $\tau_{1.5}$ and $\lambda(\theta|{\bf n})$ respectively. The maximum
among $P_{\tau_0}$, $P_{\tau_{1.5}}$ and $P_\lambda$ is defined as the {\em probability of clustering}.
\item{Consistency of clustering:} If, for a given realization of noise, the $N_{rep}$ fitness values are
 found to be clustered, then the cluster is defined to be {\em consistent} if (i) the fitness values
are also clustered for $N_p^\prime$ sufficiently greater than $N_p$, and
(ii) the absolute difference between the maximum fitness values $\rho$ and $\rho^\prime$, corresponding to
 $N_p$ and $N_p^\prime$ respectively, is $\leq 10\%$ of their mean, $(\rho+\rho^\prime)/2$.
We define the {\em consistency of clustering} as
the fraction of trials in which the clusters are consistent.
\end{description}
We deem a given combination of design variable values {\em acceptable}
 if both the probability and the consistency of clustering exceed 0.9 for that combination.
Of all the combinations that are acceptable, the optimal is chosen to be the
 one that has the lowest computational cost in terms of the mean number of template
evaluations.

%%%%%%%%%%%%%%%%%%%%%%%%%
\subsection{Simulations}
The tuning procedure described above is now applied to the two design variables $N_p$ and $N_t$.
The following set of points  is used to find the acceptable combinations:
\begin{eqnarray}
N_p &\in& \{42,81,121\}\nonumber \\
N_t &\in& \{20, 40, 80, 120, 160\}\nonumber
\end{eqnarray}
This particular domain in the $N_p$-$N_t$ plane is chosen based on our
empirical experience with PSO. ($N_p=42$ corresponds to a 7-by-6 grid of initial positions, $81$ to a 9-by-9 and $121$ to an 11-by-11 one.)
The number of trials  is 50 and
each realization of noise is 64 seconds long with sampling interval $\delta_s = 1/2048$~sec.

The tuning procedure proceeds as follows.
\paragraph{Computational cost --} For each  point in the $N_p$-$N_t$ plane,
we record  the mean number of fitness function evaluations. The
results are shown in Table~\ref{compcost_table}.
\begin{table}
\caption{\label{compcost_table}
Computational cost of PSO on data with no signals. For each combination of $N_p$ and $N_t$, the
mean number of fitness function evaluations is listed along
with the maximum (superscript) and minimum (subscript) over $50$ trials. The mean values have been rounded
off to the nearest integers.
}
\begin{ruledtabular}
\begin{tabular}{c|ccc}
& $N_p=42$ & $81$ & $121$\\\hline
$N_t=20$& $8309^{12768}_{5250}$ & $16284^{21465}_{8910}$ & $25006^{39688}_{13310}$\\
40 & $17401^{24486}_{9618}$ & $31694^{40824}_{19521}$ & $44632^{61105}_{25410}$ \\
80 & $28920^{37338}_{22302}$ & $52669^{66825}_{35559}$ & $74115^{95469}_{53119}$ \\
120 & $38567^{51450}_{32550}$ & $69982^{85293}_{49410}$ & $101495^{143990}_{75262}$\\
160 & $48147^{68880}_{38808}$ & $86759^{109755}_{68040}$ & $126346^{161535}_{98010}$
\end{tabular}
\end{ruledtabular}
\end{table}
\paragraph{Probability of clustering --}
Table~\ref{table_tuning} lists the probability of clustering for each combination of $N_t$ and $N_p$.
Note that for the combination $N_p=121$ and $N_t=40$,   $P_{\tau_0}= 94\%$
is very different from $P_\lambda=76\%$ and $P_{\tau_{1.5}}=80\%$. This suggests that
the abnormally high value of $P_{\tau_0}$ here is most likely a statistical outlier. Therefore,
we do not consider this combination as having a probability of clustering $\geq 90\%$.
\begin{table}
\caption{\label{table_tuning}
Probability of clustering  for different combinations of $N_p$ and $N_t$.
For each combination, the
fraction of trials (in \%) $P_\lambda$, $P_{\tau_0}$ and $P_{\tau_{1.5}}$
 for which the fitness, $\tau_0$ and $\tau_{1.5}$ values respectively were found to be clustered
are listed. The probability of clustering, shown in bold,
is the maximum over $P_\lambda$, $P_{\tau_0}$ and $P_{\tau_{1.5}}$. The number of trials for each
combination is 50.}
\begin{ruledtabular}
\begin{tabular}{c|ccc}
 & $N_p=42$ & $81$ & $121$ \\\hline
$N_t=20$&$\begin{array}{c}
(P_\lambda)66\\(P_{\tau_0}) {\bf 74}\\(P_{\tau_{1.5}})68
\end{array}$ &$\begin{array}{c}
60\\ {\bf 72}\\72
\end{array}$ &$\begin{array}{c}
70\\ {\bf 82} \\82
\end{array}$\\\hline
%%%%%
$40$ &$\begin{array}{c}
72\\ 82\\ {\bf 86}
\end{array}$ &$\begin{array}{c}
76\\ {\bf 88} \\76
\end{array}$ &$\begin{array}{c}
76\\ {\bf 94} \\80
\end{array}$\\\hline
%%%%%
$80$ & $\begin{array}{c}
84\\ 84\\ {\bf 88}
\end{array}$&$\begin{array}{c}
84\\ {\bf 90} \\86
\end{array}$ &$\begin{array}{c}
90\\ {\bf 92} \\92
\end{array}$ \\\hline
%%%%%
$120$ &$\begin{array}{c}
72\\ {\bf 78} \\68
\end{array}$ &$\begin{array}{c}
88\\ {\bf 92} \\88
\end{array}$ &$\begin{array}{c}
{\bf 96}\\92 \\96
\end{array}$ \\\hline
%%%%%
$160$ &$\begin{array}{c}
82\\ {\bf 88} \\78
\end{array}$ &$\begin{array}{c}
86\\ {\bf 86} \\80
\end{array}$ &$\begin{array}{c}
94\\ {\bf 94} \\92
\end{array}$

\end{tabular}
\end{ruledtabular}
\end{table}
\paragraph{Consistency of clustering --}
 Referring to Table~\ref{table_tuning}, 
we see that the consistency test is required only for $N_p\geq 81$ and $N_t\geq 80$ for which, as per the 
definition of acceptability above, the probability of clustering is $\geq 90\%$.
Further, for a given $N_t$, the computational cost is lower for $N_p=81$ than $N_p=121$.
Hence, we only tune over $N_t\geq 80$ for $N_p=81$.
No extra work is required for obtaining these results
since for each trial, the same data realization was used for both $N_p=81$ and
$N_p=121$ and the latter can be used to check if a cluster found by the former was consistent or not. In other words,
$N_p=81$ and $N_p^\prime = 121$ in the definition of the consistency of clustering given earlier.

We obtain the following results for the consistency of clustering:
  $ 91\%$,   $93\%$ and   $95\%$ for $N_t=80$, $120$ and $160$ respectively. Thus, according to our
final criterion, we pick  $N_p=81$ and $N_t=80$ as this is the acceptable combination with the lowest computing cost (c.f., Table~\ref{compcost_table}).

\subsection{Trials with no clustering}
\label{trials_noclustering}

So far, we have focussed on clustering as the main indicator of success in locating the true
global maximum. Does this imply that  in the trials in which there is no clustering, PSO fails
to locate the global maximum? To address this, we carried out the following test.
First, we retain the maximum among the $N_{rep}$ fitness values from each trial.
For each combination of $N_p$ and $N_t$, we divide the set of maximum fitness values
 into two disjoint subsets: one in which all parameters, the two chirp-times and the fitness,
 were clustered and the other in which at least
one parameter did not show clustering. For the former set, clustering of all three parameters 
is a strong indicator of successful termination near the global maximum. A two-sample
Kolmogorov-Smirnov (KS) test~\cite{wilcox} is carried out to see if
the two subsets were drawn from the same parent distribution. The
results are summarized in Table~\ref{compGoodBadClust}. 

As can be
seen from the table,  in all cases the test supports the hypothesis
that the maximum fitness value is drawn from the same distribution
irrespective of the clustering of the parameters. That this is a non-trivial
result is further supported by the fact that if the same test is done with fitness
values other than the maximum one, the null hypothesis is rejected strongly.
Table~\ref{compGoodBadClust} shows the results from the same
test but using the set of minimum fitness values. In this case, it
is seen that the values are drawn from different distributions, at
least for $N_p=81$. Thus, we conclude that even in the absence of
clustering, the PSO run that yields
 the maximum fitness value terminates, with high probability, in the vicinity of the global maximum for the $N_p=81$
and $N_t=80$ combination.

\begin{table}
\caption{
\label{compGoodBadClust}
Statistical difference in the distribution of maximum fitness values.  The
 table entries are the significance values from a
two sample Kolmogorov-Smirnov test with the null hypothesis: the
maximum fitness values from trials that show clustering of all
parameters and trials that do not are drawn from the same parent
distribution. The numbers in parentheses are the significance values
for the test done with minimum fitness values. }
\begin{ruledtabular}
\begin{tabular}{c|cc}
& $N_p=81$ & $121$\\
\hline\\
$N_t=80$ & $0.5 (2\times10^{-2})$& $ 0.9 (0.7)$ \\
120 & $0.9 (2\times 10^{-2})$ & $0.4 (0.4)$\\
160 & $0.7 (8\times10^{-4})$ & $ 0.4 (0.9)$
\end{tabular}
\end{ruledtabular}
\end{table}

 We have traced the lack of clustering to the presence of distant
peaks in the fitness function that are similar in value. The probability of this happening in the presence of
a sufficiently strong signal is very small, but this need not be so for noise-only data.

\subsection{Comments}
\label{caveats_tuning}

We have demonstrated a systematic tuning procedure  for
the design variables of PSO. It is important to note that no prior information
 about any special features of the fitness function was used.
Hence, the procedure would stay the same if the testbed were changed.

 A larger number of trials or a finer spacing of grid points in $N_p$ and $N_t$
will probably lead to a different end result. Instead of $N_p=81$, for example,
 $N_p=121$ may turn out to be the right choice. However, the main goal in this paper is to test the viability of PSO and,
for this purpose, a coarse tuning such as the one presented here is adequate.
Besides, a significant investment in refining the results of the tuning procedure would be rendered obsolete with future
improvements in the  implementation of PSO. Until a version of PSO is developed
that is hard to improve upon, a strong focus on the results from tuning,  as opposed
to improvements to the tuning procedure itself, is not of much use.

For $N_p=121$, Table~\ref{compGoodBadClust} shows that the minimum fitness values obtained with and without
clustering are also mutually consistent. This observation suggests an alternative approach to tuning where one of the
measures used for picking the optimal
combination is this type of consistency. We leave this for future work to address.

%%%%%%%%%%%%%%%%%%%%%%%%%%%%%%%%%%%%%%%%%%%%%%%%%%%%%%%%%%%%%%%%%
\section{Results with signal present}
\label{results}
\begin{figure}
\includegraphics[scale=0.6]{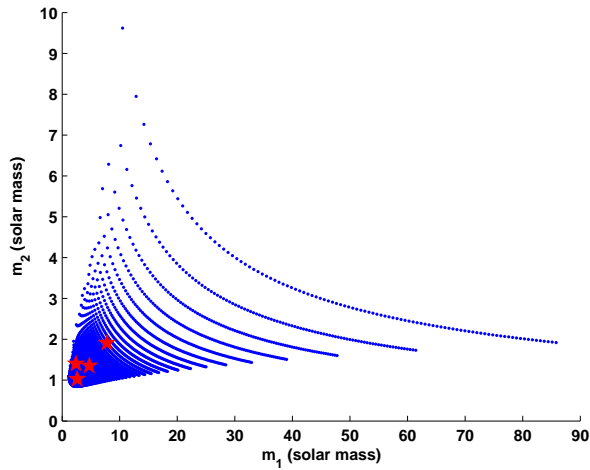}
\caption{
\label{massplane}
The region in the $m_1$, $m_2$ plane corresponding to the physically valid  part of the
$\tau_0$, $\tau_{1.5}$ plane. The region is indicated by taking a regular grid of points in the
$\tau_0$, $\tau_{1.5}$ plane and mapping them to the corresponding values of $m_1$, $m_2$,
where by convention $m_1 \geq m_2$. The $\star$ markers shows
the signal locations used in the simulations.
}
\end{figure}

In this section, we describe the results of simulations performed with signals added to data.
We quantify the performance of PSO at
four different values of signal SNR and four different locations in the $\tau_0$, $\tau_{1.5}$ plane,
\begin{eqnarray}
{\rm SNR} &\in& \{9.0, 8.0, 7.0, 6.0\}\;,\nonumber\\
(\tau_0,\tau_{1.5}) & \in & \{(5.0,0.6),(10.0,0.75),\nonumber\\
&& (16.0,0.762),(20.0,0.9)\}\;.\nonumber
\end{eqnarray}
(The units for both $\tau_0$ and $\tau_{1.5}$ are in seconds). The
corresponding  masses (in $M_\odot$)
 of the binary
components are, respectively,
$\{(m_1=7.78,m_2=1.91),(4.71,1.35),(2.40,1.40),(2.61,1.03)\}$. Fig.~\ref{massplane} shows the
physical part of the search region mapped into the
$m_1$, $m_2$ plane along with the signal locations.

 For each combination of signal location and SNR,
  50 independent data realizations
 are generated.
 The length of each realization is $64$~sec, with $\delta_s = 1/2048$~sec, and the signal is added at an offset of $10$~sec from the start.

%%%%%
\subsection{Qualitative changes induced by a signal}
\label{performance_signal}

It is instructive to
observe how a signal affects the behavior of the swarm. In general, the presence of the signal
leads to a broadening of the peak in the fitness function.
 As is well known,
the broadening is more pronounced in one direction, due to the correlation between estimation errors,
 leading to the appearance of a thin ridge-like feature (c.f., Fig~\ref{gridsnr}).

The particles begin by moving
randomly in the parameter space but each time a particle crosses the ridge, its {\em pbest} tends to fall closer to the
flanks of the ridge. As time progresses, the {\em pbest} of all particles cluster around the ridge. This increases its attractive power
in the acceleration of the particles, progressively drawing  more particles into exploration of the fitness function along the ridge.

\begin{figure}
\includegraphics[scale=0.6]{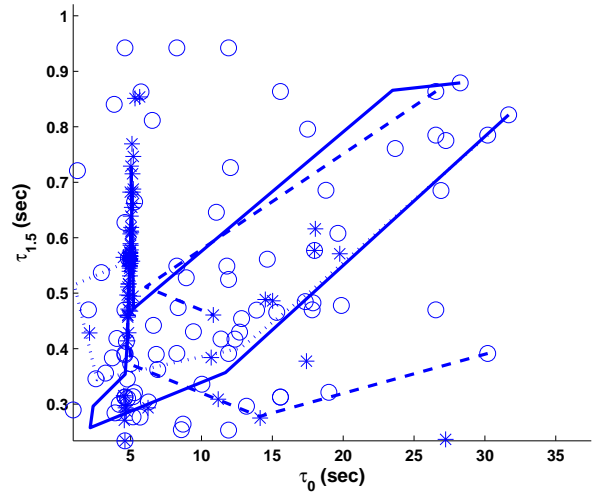}
\caption{
\label{pso_movie}
Evolution of a swarm in the presence of a signal. The `$\circ$' and `$\ast$' markers show the {\em pbest} locations of $N_p=81$ particles
when 5\% and 60\%  of the total number of steps were completed respectively. The lines show the paths followed by the {\em pbest} locations
of 5 representative particles between these two steps. With time, the {\em pbest}
locations tend to congregate around the ridge-like feature produced by a signal.
}
\end{figure}

Fig.~\ref{pso_movie} shows snapshots of PSO at different stages in the search and the progressive clustering of {\em pbest} locations
is seen clearly.
A key point to note here is that no prior knowledge is built into PSO about the ridge-like feature. It is found by the particles as
they explore the search region.

\subsection{Figures of merit}

In order to quantify the performance of PSO in the presence of a signal,
we look at two figures of merit.
The first is the probability of clustering defined in Sec.~\ref{tuning_fom}. Since
the tuning procedure requires a minimum value of 90\%, the probability of clustering
in the presence of a strong signal should be significantly higher but it should be consistent with the pure
noise case for weak signals.

When a signal is added to the data, we do not need the consistency  of clustering criterion
of Sec.~\ref{tuning_fom} in order to confirm the association of a cluster
with the global maximum. Since we know the location of the signal and since the expectation
of the fitness function must be maximum at that location, it suffices to check if the maximum
fitness in the cluster is larger than the value at the signal location. Our second figure of merit,
therefore, is the fraction of trials in which this occurs.  Ideally, this figure of merit should be unity.

Table~\ref{summary_results} reports the first figure of merit  for each
combination of signal SNR and location.  As expected, for the case of
strong signals (SNR$\geq 7$) the probability of clustering is always,
and often significantly, higher than 90\%. For the weak signal SNR of $6.0$, the probability of clustering
has an average value of 91\% which is statistically consistent with the pure noise case of 90\%.

\begin{table}
\caption{\label{summary_results}
Probability of clustering for simulations with signal present in the data.  For each combination of signal SNR and location, 
the fraction of trials (in \%) $P_\lambda$, $P_{\tau_0}$ and $P_{\tau_{1.5}}$
 for which the fitness, $\tau_0$ and $\tau_{1.5}$ values respectively were found to be clustered
are listed. The probability of clustering, shown in bold,
is the maximum over $P_\lambda$, $P_{\tau_0}$ and $P_{\tau_{1.5}}$. The number of trials for each
combination is 50.
}
\begin{ruledtabular}
\begin{tabular}{c|cccc}
& $\begin{array}{c}
(\tau_0,\tau_{1.5})\\
=(5.0,0.6)
\end{array}$ & $(10.0,0.75)$ & $(16.0,0.762)$ & $(20.0,0.9)$\\
\hline
SNR=9.0 &$\begin{array}{c}(P_\lambda){\bf 98}\\(P_{\tau_0})94\\(P_{\tau_{1.5}})94\end{array}$ &$\begin{array}{c}{\bf 94}\\92\\94\end{array}$ & $\begin{array}{c}94\\{\bf 96}\\94 \end{array}$& $\begin{array}{c}90\\{\bf 98}\\94\end{array}$\\\hline
8.0 &$\begin{array}{c}96\\{\bf 96}\\92\end{array}$ &$\begin{array}{c}{\bf 98}\\96\\96\end{array}$ &$\begin{array}{c}{\bf 98}\\92\\92\end{array}$ &$\begin{array}{c}94\\92\\{\bf 96}\end{array}$ \\\hline
7.0 &$\begin{array}{c}{\bf 96}\\96\\90\end{array}$ & $\begin{array}{c}{\bf 92}\\90\\88\end{array}$& $\begin{array}{c}{\bf 98}\\94\\98\end{array}$& $\begin{array}{c}{\bf 92}\\88\\90
\end{array}$\\\hline
6.0 &$\begin{array}{c}92\\{\bf 94}\\84\end{array}$ &$\begin{array}{c}82\\{\bf 86}\\86\end{array}$ &$\begin{array}{c}{\bf 88}\\86\\84\end{array}$ &$\begin{array}{c}94\\{\bf 96}\\96
\end{array}$
\end{tabular}
\end{ruledtabular}
\end{table}

\begin{figure}
\includegraphics[scale=0.6]{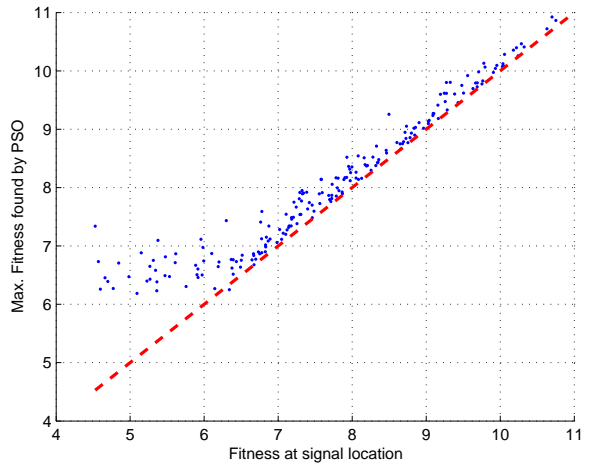}
\caption{\label{snr_scatterplot}
Scatterplot of maximum fitness value found by PSO (Y axis) against the
 value at the known signal location, $\tau_0=10.0$~sec, $\tau_{1.5}=0.75$~sec, for
all signal SNR values. In one trial, the maximum fitness value (near 6.0 on the Y axis) dips below the
line of equality (dashed).
}
\end{figure}

As far as the second figure of merit is concerned, we find that it is unity for all combinations of
signal SNR and locations except for one,
namely, $SNR=8.0$, $\tau_0=10.0$~sec and $\tau_{1.5}=0.75$~sec, for which it was 0.98.
Fig.~\ref{snr_scatterplot} shows  the scatterplot between the maximum fitness found by PSO and the value at this signal location
for all signal SNR values. It is seen that in one trial the maximum fitness fell below the value at the signal location. However,
the two values are so close that the figure of merit should be considered to be practically unity for this case too.

Taken together, the figures of merit show that
PSO almost always terminates near the true global maximum when a sufficiently strong signal is present. When the signal is weak, we recover
the performance ensured by the tuning procedure for the pure noise case.

%%%%%%
\subsection{Signal detection and parameter estimation}

In order to cast the results obtained so far in terms of signal detection and parameter estimation
performance, we choose the maximum fitness value found
 over the $N_{rep}$ runs as the detection statistic and the location corresponding to the maximum fitness value as
the estimator for the chirp time parameters.

\subsubsection{Detection}
It was discussed in Sec~\ref{trials_noclustering}
that, after tuning PSO, the distribution of the detection statistic in trials with and without
clustering remains the same. As the simultaneous clustering of the two chirp-times and the fitness values 
indicates termination in the vicinity of the global maximum,
it follows that the probability distribution of
the PSO detection statistic is about the same as that of the global maximum.
 Strictly speaking, the PSO detection statistic will always have a value less than
the global maximum but, given our termination criterion, the relative difference between the two is less than
3\%. Thus, the false alarm probabilities, for a given detection threshold, corresponding to the PSO detection
statistic and the true global maximum  are also
nearly the same, with the former being slightly smaller.

 In the presence of signals with an SNR of 8 or higher, which is the typical
value sought in a real detection, it was shown that almost all trials exhibit
 clustering and that the detection statistic value was always higher than the fitness at the true signal
location. Hence, the distribution of the detection statistic in the presence of a signal also closely follows that of
the true global maximum. Strictly speaking, as with the false alarm probability, the detection probability will be slightly smaller for
the PSO detection statistic, for a given threshold, as compared to that for the true global maximum.

The above line of reasoning suggests that the Receiver Operating Characteristics (ROC) of
the PSO detection statistic should nearly be the same as that of the true global maximum. The only way to rigorously verify this
is to carry out simulations with a large number of trials in which both PSO and grid-based searches are performed. This
is a computationally expensive task which we plan to undertake in the future.
However, it is important to note that such a comparison may not be possible
for searches that are too expensive for a grid-based search.

\subsubsection{Estimation}
Table~\ref{param_estimation} summarizes
the errors in the estimation of the chirp time parameters for signal SNR values $\geq 8$ at
the different signal locations used in the simulations. Each entry in the table is an estimate of
 the root mean-square error (rmse) defined as,
\begin{equation}
{\rm rmse}(\theta) = \left[E\left[\left( \widehat{\theta}-\theta\right)^2\right]\right]^{1/2}\;,
\end{equation}
where $\widehat{\theta}$ is the estimator of $\theta$.  The rmse includes the effects of both estimator variance and bias.

\begin{figure}
\includegraphics[scale=0.6]{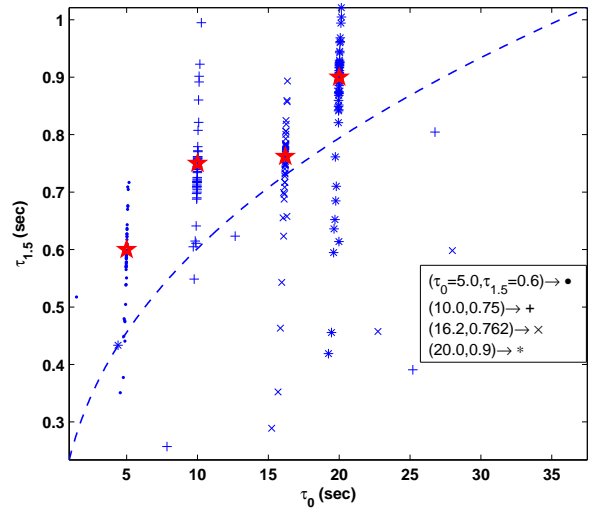}
\caption{
\label{snr_8_paramest}
Estimation of parameter values for a signal SNR of 8.0.  The true locations of the signals are indicated by the $\star$ marker and each
of the markers, $\bullet$, $+$, $\ast$ and $\times$, indicates an estimated location corresponding
to one of the true locations. The association between the markers and the true
signal locations is indicated in the figure. For each true signal location, the simulation consisted of 50 trials.
}
\end{figure}

Since the search region in the current testbed includes unphysical chirp time parameters,
 the global maximum and, hence, the estimated chirp times
fall there in some trials.  Fig.~\ref{snr_8_paramest} shows an example
where the estimates from all the trials are shown for a signal SNR of 8.
In an improved implementation of PSO,
blocking the unphysical region should improve parameter estimation accuracy significantly. As an indicator of this, we also
show in Table~\ref{param_estimation}, the rmse obtained by dropping the trials where the estimate fell in the unphysical
region. It is seen that the errors are reduced significantly, especially for the lower signal SNR value for which there is more
 scatter into the the unphysical region.

A comparison of Table~\ref{param_estimation} with existing results~\cite{balasubramanian96}
shows that the estimation errors due to PSO are consistent with grid-based searches.

\begin{table}
\caption{
\label{param_estimation}
Signal parameter estimation errors with PSO. Each entry in the table is of the form $a(b)$, where $a$ and $b$ are the estimated
root mean square errors (rmse) for $\tau_0$ and $\tau_{1.5}$ respectively
 (expressed as a percentage of the true parameter value).
  In each row, the top and bottom pairs of numbers refer to mse obtained without and with the
 physical boundary cut respectively. All the numbers have been rounded off, given the expected precision
from the 50 trials used per combination of signal  SNR and location.
}
\begin{ruledtabular}
\begin{tabular}{c|cccc}
& $\begin{array}{c}
(\tau_0,\tau_{1.5})\\
=(5.0,0.6)
\end{array}$& $(10.0,0.75)$ & $(16.2,0.762)$ & $(20.0,0.9)$ \\
\hline
SNR=9.0 &
        $\begin{array}{c}2(13)\\2(12) \end{array}$&
             $\begin{array}{c}1(11)\\0.5(6) \end{array}$
                 &$\begin{array}{c}0.3(6.0)\\0.2(4) \end{array}$
                     &$\begin{array}{c}1(11)\\0.2 (4) \end{array}$\\
\hline
8.0 &
       $\begin{array}{c} 44(13)\\10.5(10) \end{array}$
         &$\begin{array}{c}32(16) \\1(10)\end{array}$
           & $\begin{array}{c}12(16)\\0.3(5) \end{array}$&
                  $\begin{array}{c}11(17)\\12(10)\end{array}$\\
\end{tabular}
\end{ruledtabular}
\end{table}

 %%%%%%%
\subsection{Computational cost}
\label{comparison_compcost}
The number of fitness function evaluations for each combination of signal SNR and location are
shown in Table.~\ref{compcost_signal}.
It is seen  that for a signal SNR of 9.0, the maximum number of evaluations
 is about the same as the mean in the pure noise case (c.f., Table~\ref{compcost_table}).
This reduction is consistent with the fact that a strong signal makes  it easier
for the swarm to find the global maximum.
\begin{table}
\caption{\label{compcost_signal}
Computational cost of PSO on data containing a signal.
 For each combination of signal SNR and location, the
mean number of fitness function evaluations, over $50$ trials, is listed along
with the maximum (superscript) and minimum (subscript). All
numbers are in units of $10^4$ and rounded off to a single digit of precision.
}
\begin{ruledtabular}
\begin{tabular}{c|cccc}
& $\begin{array}{c}
(\tau_0,\tau_{1.5})\\
=(5.0,0.6)
\end{array}$ & $(10.0,0.75)$ & $(16.0,0.762)$ & $(20.0,0.9)$\\\hline
SNR=9.0 & $4.4_{3.8}^{5.2}$ & $4.7_{4.1}^{5.7}$ &  $4.8_{4.2}^{5.6}$ &
$4.7_{4.2}^{5.8}$ \\
8.0 &  $4.5_{3.8}^{6.0}$ & $4.7_{3.7}^{6.5}$ & $4.8_{4.1}^{5.6}$ &  $4.8_{4.2}^{5.9}$ \\
7.0 &  $4.7_{3.6}^{6.4}$ &  $4.8_{3.9}^{6.3}$ &  $4.9_{3.7}^{6.2}$ &
$4.8_{3.8}^{5.9}$ \\
6.0 &  $4.5_{3.3}^{6.1}$ &  $4.8_{2.9}^{7.0}$ &  $4.7_{3.2}^{5.9}$ & $4.7_{3.6}^{6.0}$ \\
\end{tabular}
\end{ruledtabular}
\end{table}

For ground-based detectors, the dominant computational cost  comes from  the pure noise case.
Although our tuning procedure produced $N_p=81$ and $N_t=80$
as the optimal combination, there is statistical uncertainty in this result due to the
finite and somewhat small number of trials used.  To make our estimate of the computational cost conservative,
we use the combination $N_p=121$ and $N_t=80$ instead for which
all the performance measures are significantly better. From Table~\ref{compcost_table},
 the typical number of fitness evaluations required for the testbed considered here
is $\sim 7\times 10^4$ with a spread of about $\pm 2\times10^4$.
Of this, the termination criterion itself accounts for a fixed number, $(N_t=80)\times (N_p=121)=9680$, of
 evaluations.

A grid-based search provides a convenient perspective for evaluating the computational cost of PSO.
According to~\citep{owen99}, for 2PN waveform and
initial LIGO noise PSD, the number of fitness function evaluations required in a single grid with
a minimal match of
0.97 ($\Rightarrow \alpha = 0.03$) is $1.1\times10^4$  if the minimum mass used for constructing the
 template waveforms is $1M_\odot$.
In the current testbed, the search region in the mass plane (c.f., Fig.~\ref{massplane}) is not
the simple one considered in~\cite{owen99} although the minimum masses are similar. Additionally,
 \citep{owen99} uses an
analytic fit for the noise PSD that differs from the one used here.
Ignoring these differences we find that the current implementation of PSO
requires about 7 times as many evaluations, on the average, as a grid-based method.

%%%%%%%%%%%%%%%%%%%%%%%%%%%%%%%%%%%%%%%%%%%%%%%%%%%%
\section{Conclusions}
We applied PSO to the binary inspiral testbed where the main challenge was to locate the global
maximum of a highly multi-modal fitness function. Such functions,
with an unpredictable number of extrema having random locations and sizes, are typical in
GW data analysis. 

In response to the questions posed at the beginning,
the results obtained from simulations show clearly that:
\begin{enumerate}
\item PSO is a viable method
for signal detection and estimation in GW data analysis as it can successfully handle
the challenge of high multi-modality presented by such problems.
\item Good performance was achieved by tuning only two out of the nine design variables involved in the method.
Thus, PSO is a stochastic method that offers the possibility of having a small number of design variables in practice.
\item The design variables were tuned using a systematic procedure that does not require
 any prior information about features of the fitness function. As such, the procedure should be widely applicable 
to other stochastic methods also.
\item PSO is about 7 times more expensive than a grid-based search in the number
of fitness function evaluations required.
\end{enumerate}
The higher cost of PSO is not surprising since grid-based searches are usually more efficient than stochastic methods in low-dimensional
 problems such as the one considered here. The performance gain of stochastic methods appears due to the slower rise in 
their computational cost, with increase in dimensionality, compared to the exponential one of grid-based searches. Therefore, 
we expect PSO to be cheaper than grid-based searches in higher dimensional problems. However,
a definitive answer requires an actual test on problems 
 such as the inspiral of high mass spinning binary components or the LISA Galactic Binary resolution problem~\cite{babak+etal:mldc2}.
The demonstration in this paper
 that PSO can handle the more serious challenge of high multi-modality is the first step towards such
future investigations.

The computational cost of PSO may be significantly reduced by taking into account the physical boundary in parameter 
space (see Fig.~\ref{snr_8_paramest}). 
The current implementation of PSO requires the
search to extend over a large unphysical region. In fact, 
as far as the binary inspiral problem goes, we find this to be 
the most outstanding issue.
We have tried the invisible walls condition with the curved physical boundary but find that the
performance of PSO is negatively affected.
Specifically, termination takes a much longer time and the probability
of clustering is significantly reduced.
This behavior is attributable to the curved shape of the boundary allowing a
significantly larger number of particles to escape the search domain. Once outside,
 particles contribute nothing to the search and keep moving until they are pulled back.

To solve this problem, it appears inevitable that the dynamical equations of PSO must
be modified. For signals other than binary inspirals, such as
Galactic Binaries in the case of LISA, the nature of the boundary problem
 would be different and it may not be an issue in some applications.

Finally, a comment about the use of Gaussian, stationary noise in
the testbed. We emphasize here that PSO is a method for finding the
global maximum of a fitness function {\em irrespective} of what
produces that peak, a genuine GW signal or an instrumental
transient. Since, the implementation of PSO in this paper uses no
prior information about features of the fitness function, it should
find the peak regardless of its source. Thus, there should be no
significant difference in the performance of PSO and a grid-based
method even for non-stationary, non-Gaussian noise. In future work,
we will verify this explicitly by using non-GW signals in our
simulations.

\label{conclusions}
%%%%%%%%%%
\begin{acknowledgements}
This work was supported by Research Corporation award CC6584.  YW
was supported by NSFC grant 10773005 of China and NSF award
HRD-0734800 to the Centre for Gravitational Wave Astronomy at the
University of Texas at Brownsville.
\end{acknowledgements}

%%%%%%%%%%%%%%%%%%%%%%%%%%%%%%%%%%%%%%%%%%%%%%%%%%%%%%%%%%%%%%%%
\appendix
\section{Details of the signal waveform}
\label{details_signal}
\subsection{Chirp-time parameters}
The chirp-time parameters, $\{\tau_a\}$, $a=0,1,1.5,2$, are given in
terms of the masses, $m_1$ and $m_2\leq m_1$, of the binary components (we use $c=G=1$),
\begin{eqnarray}
\tau_0&=&\frac{5}{256}M^{-5/3}\eta^{-1}(\pi f_a)^{-8/3}\;,\\
\tau_1&=&\frac{5}{192\mu (\pi
f_a)^{2}}\left(\frac{743}{336}+\frac{11}{4}\eta \right)\;,\\
\tau_{1.5}&=&\frac{1}{8\mu}\left(\frac{M}{\pi^2{f_a}^5}\right)^{1/3}\;,\\
\tau_2&=&\frac{5}{128\mu}\left(\frac{M}{\pi^2{f_a}^2}\right)^{2/3}
\left(\frac{3058673}{1016064}+\frac{5429}{1008}\eta+\right.\nonumber\\&&\left. \frac{617}{144}\eta^2\right)\;,
\end{eqnarray}
where $M=m_1+m_2$ is the total mass of the compact binary, $\mu=m_1m_2/M$ is the
reduced mass and $\eta=\mu/M$.

Since all the chirp time parameters depend on $m_1$ and $m_2$, only two of them are independent.
It is convenient to choose $\tau_0$ and $\tau_{1.5}$ as the independent parameters since
 $M$ and $\mu$ can be obtained algebraically from them,
\begin{eqnarray}
\mu&=&\frac{1}{16f_a^2}\left(\frac{5}{4\pi^4\tau_0\tau_{1.5}^2}\right)^{1/3}\;,\\
M&=&\frac{5}{32f_a}\frac{\tau_{1.5}}{\pi^2\tau_0}\;,
\end{eqnarray}
allowing $\tau_{1}$ and $\tau_2$ to be obtained algebraically from $\tau_0$ and $\tau_{1.5}$.

%%%%%%%%%
\subsection{The phase function}
In Eq.~\ref{stat_phase_eqn}, the function $\psi(f;\theta)$ is given by,
\begin{eqnarray}
\psi(f;\theta)&=&
\sum_{i\in \{0,1,1.5,2\}}\!\!\!\alpha_i(f)\tau_i\;, \\
\alpha_0(f)&= & 2\pi f - \frac{16\pi f_a}{5} +\frac{6\pi
f_a}{5}\left(\frac{f}{f_a}\right)^{-5/3}
\;,\\
\alpha_{1}(f)&= &  2\pi f - 4\pi f_a + 2\pi
f_a\left(\frac{f}{f_a}\right)^{-1}\;, \\
\alpha_{1.5}(f)&= & - 2\pi f + 5\pi f_a - 3\pi
f_a\left(\frac{f}{f_a}\right)^{-2/3}
\;, \\
\alpha_{2}(f)&= & 2 \pi f - 8\pi f_a + 6\pi
f_a\left(\frac{f}{f_a}\right)^{-1/3}\;.
\end{eqnarray}
The functions $\alpha_a$, $a=0,1,1.5,2$ and the $f^{-7/6}$
factor in the amplitude of the signal (Eq.~\ref{stat_phase_eqn})
 can be pre-computed and stored, reducing
the computational cost of generating waveforms.

%%%%%%%%%%%%%%%%%%%%%%%%%%%%%%%%%%%%%%%%%
\end{document}